\documentclass[prl,twocolumn,superscriptaddress,showpacs,floats,floatfix]{revtex4}
\usepackage{epsfig}

\begin{document}

\preprint{1}

\title{Observation of Magnetic Edge State and Dangling Bond State on Nanographene 
in Activated Carbon Fibers}

\author{Manabu Kiguchi}
\affiliation{Department of Chemistry, Tokyo Institute of Technology, 2-12-1 Ookayama, Meguro-ku, Tokyo 152-8551, Japan.}
\author{Kazuyuki Takai}
\affiliation{Department of Chemistry, Tokyo Institute of Technology, 2-12-1 Ookayama, Meguro-ku, Tokyo 152-8551, Japan.}
\author{V. L. Joseph Joly}
\affiliation{Department of Chemistry, Tokyo Institute of Technology, 2-12-1 Ookayama, Meguro-ku, Tokyo 152-8551, Japan.}
\author{Toshiaki Enoki}
\affiliation{Department of Chemistry, Tokyo Institute of Technology, 2-12-1 Ookayama, Meguro-ku, Tokyo 152-8551, Japan.}
\author{Ryohei Sumii}
\affiliation{Institute of Materials Structure Science, High Energy Accelerator Research
Organization, Tsukuba, Ibaraki 305-0801, Japan.}
\author{Kenta Amemiya}
\affiliation{Institute of Materials Structure Science, High Energy Accelerator Research
Organization, Tsukuba, Ibaraki 305-0801, Japan.}

\date{\today}
\begin{abstract}
The electronic structure of nanographene in pristine and fluorinated activated carbon fibers 
(ACFs) have been investigated with near-edge x-ray absorption fine structure (NEXAFS) and 
compared with magnetic properties we reported on previously. In pristine ACFs in which magnetic 
properties are governed by non-bonding edge states of the $\pi$-electron, a pre-peak assigned to the 
edge state was observed below the conduction electron $\pi^{\ast}$ peak close to the Fermi level in 
NEXAFS. Via the fluorination of the ACFs, an extra peak, which was assigned to the $\sigma$-dangling 
bond state, was observed between the pre-peak of the edge state and the $\pi^{\ast}$ peak in the NEXAFS 
profile. The intensities of the extra peak correlate closely with the spin concentration created upon 
fluorination. The combination of the NEXAFS and magnetic measurement results confirms the 
coexistence of the magnetic edge states of $\pi$-electrons and dangling bond states of $\sigma$-electrons on 
fluorinated nanographene sheets.
\end{abstract}



\maketitle

\section{INTRODUCTION}
\label{sec1}

The magnetism of carbon-based materials has been one of the most important and challenging 
issues in carbon science and molecule magnetism since 1950s \cite{phi1954}, and it is still a matter of 
controversy due to difficulty in obtaining well characterized materials \cite{carbon}. Recent works on 
nanographene and graphene edges have shed light on this long lasting problem and have proposed 
an edge state as a promising candidate explaining the magnetic origin in $\pi$-electron-based carbon 
materials \cite{rev2009,sol2009}. The circumference of an arbitrary shaped piece of nanograpehene is described in 
terms of a combination of zigzag and armchair edges. According to theoretical studies, edge states 
of nonbonding $\pi$-electron origin are created in the zigzag edge region even if all the edge carbon 
atoms are terminated by foreign atoms, in spite of the absence of such states in the armchair edges 
\cite{jpsj1996,jpsj1998}. Interestingly, the edge states are predicted to be strongly spin polarized. With these theoretical 
predictions as the impetus, many experimental works have been devoted to investigate the 
electronic structure and magnetic properties of nanographene inherent to the edge state by utilizing 
various techniques. These include scanning tunneling microscopy/spectroscopy (STM/STS), near 
edge x-ray absorption fine structure (NEXAFS), energy loss near edge fine structure (ELNES), 
magnetic susceptibility, electron spin resonance (ESR), magnetic force microscopy, etc 
\cite{prb2005,prb2010,prb2006,apl2006,nat2010}. 
STM/STS observations on graphene edges existing at a graphite step edge have demonstrated the 
existence of a finite local density of states assigned to the edge-state at the Fermi level (Dirac point) 
along the zigzag edge \cite{prb2005,prb2010,prb2006}. The presence of the edge-state around the Fermi level has also been 
suggested for nanographene grown on Pt(111) by carbon K-edge NEXAFS \cite{apl2006}. Recently, Seunaga 
succeeded to observe the atomically resolved electronic structure of the edge atom of graphene by 
ELNES \cite{nat2010}. From the magnetism point of view, magnetic susceptibility and electron spin 
resonance (ESR) measurements on nanoporous carbon consisting of nanographene sheets have 
demonstrated the detailed behavior of edge-state spins contributing to the magnetism of 
nanographene \cite{sol2009}. However, most of these experimental studies investigated either the electronic or 
the magnetic state only, without showing direct evidence that the edge state is responsible for the 
origin of the magnetism. Accordingly, it is an important requisite to investigate nanographene 
samples with a combination of electronic and magnetic techniques. Recently, we reported the 
existence of the edge states in nanographene and confirmed the magnetic feature of the edge state 
by investigating a graphene nanoribbons sample using NEXAFS and ESR \cite{prb2010j}. However, the issue 
of the relationship between the edge state and its magnetic feature still remains not solved due to the 
presence of trace-level magnetic impurities mentioned in our previous report.

In addition to the edge state of $\pi$-electron origin, there is another possible mechanism by 
which localized spins can be created; that is, the localized spins of $\sigma$-dangling bonds, which are 
believed to be the origin of carbon magnetism from the early stages of research on carbon 
magnetism \cite{phi1954}. For example, when the edge carbon atoms are unterminated, the $\sigma$-dangling bonds 
of these carbon atoms have localized spins. $\sigma$-dangling bonds can also be created by locally 
destroying the flat sp$^{2}$/$\pi$ hexagon network of a graphene sheet \cite{prb2007}. This can be done by bonding a 
foreign atom such as hydrogen or fluorine to a carbon atom in the interior of a graphene sheet. 
Indeed, for the latter case, the creation of localized spins was observed for highly fluorinated ACFs 
by magnetic measurements \cite{sol2009,jpsj2001}. The appearance of the localized spins in the above case was 
explained by dangling bond formation due to the local destruction of the $\pi$-conjugation at the 
interior carbon sites via fluorination. Here, again, only the magnetic properties of the fluorinated 
ACFs have been investigated with the electronic structure remaining poorly understood. 

In the present study, we discuss the electronic structure and magnetic properties of the edge 
and $\sigma$-dangling bond in nanographene in pristine and fluorinated ACFs on the basis of a 
combination of electronic (NEXAFS) and magnetic techniques (SQUID) we reported on previously 
\cite{jpsj2001}. ACF is a nanoporous piece of carbon consisting of a three dimensional disordered network of 
nanographite domains, each of which is a stack of 3-4 nanographene sheets with a mean in-plane 
size of about 2-3 nm. ACFs have a high density of edge carbon atoms (the number ratio of edge 
carbon atoms to the interior carbon atoms is ca. 30/200) due to the small size of nanographene 
sheets. ACFs are, thus, a good model system for investigating the magnetic edge state and 
$\sigma$-dangling bonds.

\section{EXPERIMENTAL}
\label{sec2}
Pristine ACFs were commercially available samples (Kurary Chemicals, FR-20; specific 
surface areas of 2000 $m^{2}/g$) prepared by the activation of phenol-based precursor materials. 
Previous studies confirmed that magnetic impurities were well below the detection level in SQUID 
and ESR measurements \cite{jpsj2001,nex}. Fluorination was carried out by the direct reaction between 
fluorine gas and ACFs. The carbon K-edge NEXAFS was measured at the soft x-ray beam line 
BL-7A in the Photon Factory in the Institute of Materials Structure Science. The ground powder 
sample was mounted on a Ta plate and loaded into the chamber maintained in ultrahigh-vacuum 
(10$^{-7}$ Pa).  NEXAFS spectra were then obtained by measuring the sample photocurrent. NEXAFS 
measurements were performed for the samples, which were characterized by X-ray photo emission 
spectra (XPS) and SQUID in our previous study \cite{jpsj2001}. The experimental details are described in this 
paper. Briefly, the degree of fluorination F/C, which represents the atomic composition ratio, was 
determined by the ratio of the C1s peak area to the F1s peak area in XPS spectra. Magnetic 
susceptibility measurement were carried out by a SQUID magnetometer (Quantum Design 
MPMS5) in a 1 T field between 2K and 380 K, where we used about 20 mg samples vacuum sealed 
in quartz tubes. 

\section{MAGNETIC EDGE STATE ON ACFS}
\label{sec2}

\begin{figure}[!t]
\epsfig{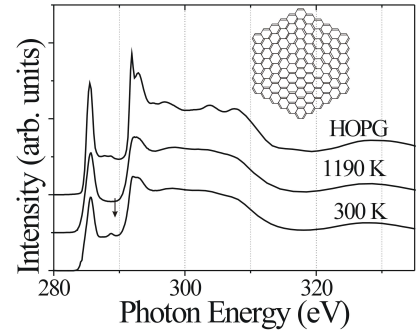}
\caption{ 
Carbon K-edge NEXAFS spectra of the ACF at room temperature, ACF heated at 1190 K, 
and HOPG. The arrow indicates the position of additional peak (see text). Inset: the schematic 
structure model of nanographene sheet consisting of 216 carbon atoms in ACF.}
\label{Fig:1}
\end{figure}

\begin{figure}
\epsfig{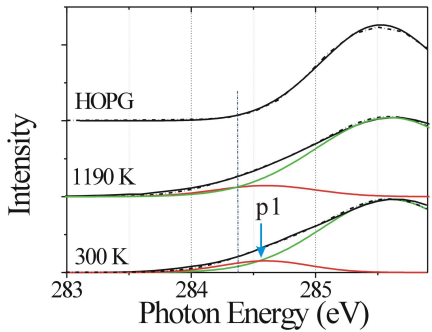}
  \caption{(color online)
The close-up of the pre-edge region of the NEXAFS spectra (line) and 
convoluted curve fit (dot-dash-line). The deconvolution comprises two Gaussian peaks 
corresponding to the edge state (p1: red curve) and $\pi^{\ast}$ state (green curve). The vertical dot-dash 
line indicates the Fermi level of HOPG (284.36 eV) (see Ref. 14.)}
\label{Fig:2}
\end{figure}

Figure~\ref{Fig:1} shows the C K-edge NEXAFS spectra of the ACFs at room temperature and ACFs 
heated at 1190 K, in addition to the spectrum of HOPG. The NEXAFS spectra were normalized with 
respect to the value of the edge jump at 340 eV, where the intensity was proportional to the amount 
of carbon atoms. All of the samples showed two peaks at 285.5 eV and 291.9 eV, which correspond 
to the C 1s to $\pi^{\ast}$ and the C 1s to $\sigma^{\ast}$ transitions, respectively. An additional small feature was 
observed at 288.5 eV for the ACFs, which can be attributed to the C-OH, C-OOH and C-H groups 
of the foreign species bonded to ACFs \cite{nex}. The feature disappeared for the ACFs heated at 1190 K. 
Figure~\ref{Fig:2} is the close-up of the pre-edge region of the NEXAFS spectra. We can see a tailing on the 
low energy side of the $\pi^{\ast}$ peak, which suggests the presence of an additional peak (p1) around the 
Fermi level. The spectral analysis was carried out for the NEXAFS spectra. First, the peak width of 
the $\pi^{\ast}$ peak was determined to be 1.5 eV from the spectrum of HOPG. Next, the spectra of the ACFs 
were fitted with two Gaussian peaks assuming that the energy and peak width of the $\pi^{\ast}$ peak were 
the same as those of HOPG. The energy and peak width of the additional peak (p1) was estimated as 
284.5eV and 0.82 eV, respectively, from the fitting results shown in Fig.~\ref{Fig:2}. The integrated intensity 
of the p1 peak was 11 $\%$ and 9 $\%$ of that of the $\pi^{\ast}$ peak for the ACFs and ACFs heated at 1190 K, 
respectively. The p1 peak survived after heating at 1190 K, although the intensity of the p1 peak 
slightly decreased. 

Usually, the graphene edge is terminated by oxygen-containing functional groups such as 
carboxyl (-COOH), carbonyl (=CO), hydroxyl (-OH), and phenol (-C$_{6}$H$_{5}$OH) groups for the ACFs 
sample exposed to the air at room temperature \cite{jpc2010}. By heating the ACFs above 1190 K in UHV, 
these groups are completely decomposed, as evidenced by the negligible content of oxygen in the 
XPS spectra \cite{XPS}. The edge would be terminated by oxygen-free functional groups such as 
hydrogen. Therefore, the existence of the p1 peak for the ACFs heated at 1190 K confirmed that the 
p1 peak originates from the intrinsic electronic state of the nanographene close to the Fermi level, 
that is, edge states in the nanographene sheet. Our recent experimental results with nanographene 
showed the decrease in intensity of the p1 peak with the heat treatment temperature above 1800 K 
\cite{XPS}. Heat treatment at temperatures above the graphitization temperature (above 1800 K) makes 
nanographene sheets fuse with each other, resulting in the successive disappearance of the 
nanographene edge \cite{sol2009}. The decrease in the p1 intensity with temperature also confirmed the 
assignment of the p1 peak to the edge state. The decrease in the intensity of the p1 peak upon 
annealing at 1190K spectra can be explained by the interaction between nanographene and 
oxygen-containing functional groups, which bond to the graphene edge. The oxygen-containing 
functional groups that are abundant in the sample at 300 K act as electron acceptors, resulting in the 
downshift of the Fermi level from the Dirac point \cite{jpc2010}. Since the center of the edge state is just at 
the Dirac point in neutral nanographene, the downshift of the Fermi level leads to an increase in the 
density of unoccupied states, which is detected by NEXAFS. By heating the ACFs above 1190 K in 
UHV, the oxygen-containing functional groups are decomposed completely and the Fermi level 
shifts back \cite{jpc2010}. Accordingly, the density of unoccupied states, that is, the p1 peak decreased by 
heating to 1190 K.

According to a previous Raman study, the size of our nanographene was estimated as $\sim$2.5 nm 
\cite{jpsj2001}. This gives information on the composition ratio of the edge carbon atoms to the carbon atoms 
in the interior of a nanographene sheet. In order to allow us to get a concrete idea of nanographene 
sheet, a representative structure of a nanographene sheet with a mean size of $\sim$2.5 nm is shown in 
the inset of Fig.~\ref{Fig:1}. It is a hexagonal shaped nanographene sheet consisting of 216 carbon atoms. The 
number of edge carbon atoms connected with two carbon neighbors is given to be 36, which 
corresponds to 17 $\%$ of all of the carbon atoms involved. The estimated contribution of the edge 
carbon atoms is in good agreement with $\sim$10 $\%$ contribution of the observed p1 integrated peak 
intensity \cite{ass}. Meanwhile, our previous magnetic susceptibility and ESR measurements of the 
ACFs samples employed in the present NEXAFS observations revealed the presence of the 
localized spins which interact with each other through an exchange interaction \cite{sol2009,jpsj2001}. Accordingly, 
the combination of the NEXAFS and magnetic measurement results confirms the existence of the 
magnetic edge state for nanographene.

\section{MAGNETIC EDGE AND DANGLING BOND STATE ON FLUORINATED ACFS}
\label{sec4}

\begin{figure}
\epsfig{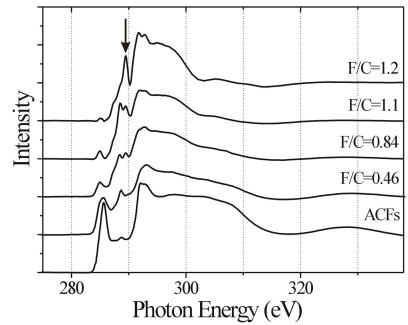}
  \caption{
The carbon K-edge NEXAFS spectra of the fluorinated ACFs at various fluorine 
concentrations. The blue arrow indicates a peak appearing at 290 eV. }
\label{Fig:3}
\end{figure}

\begin{figure}
\epsfig{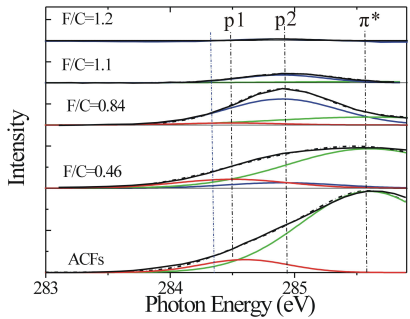}
  \caption{(color online)
The close-up of the pre-edge region of the NEXAFS spectra (black curve) 
and convoluted curve fit (dot-dash curve) of the fluorinated ACFs at various fluorine concentrations. 
The deconvolution comprises three Gaussian peaks corresponding edge state (p1: red curve), 
dangling bond state (p2: blue curve) and $\pi^{\ast}$ state (green curve). The vertical dot-dash line 
indicates the Fermi level of HOPG (284.36 eV). }
\label{Fig:4}
\end{figure}

\begin{figure}
\epsfig{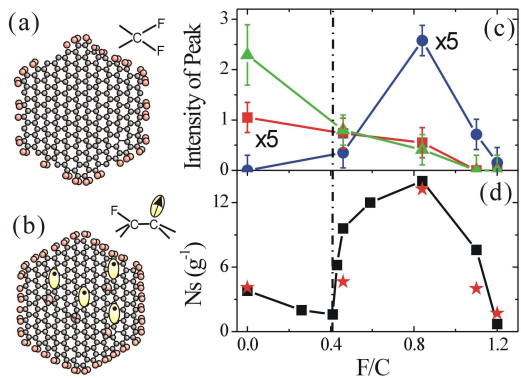}
  \caption{(color online) Schematic model of the fluorinated ACFs at (a) F/C$<$0.4, (b) F/C=0.4$\sim$0.8. 
The edge carbon atoms bonded to two neighboring carbon atoms are terminated by two fluorine 
atoms (red circles). A $\sigma$-dangling bond (ellipsoids with a dot inside) is created at a carbon site 
bonded to the carbon atom attacked by a fluorine atom in the interior of a nanographene sheet. (c) 
The intensities of the $\pi$-edge state (red squares), $\sigma$-dangling bond state (blue circles), and $\pi^{\ast}$ state 
(green triangles) peaks as a function of fluorine concentration. (d) The total localized spin 
concentration (black squares) as a function of fluorine concentration \cite{jpsj2001}. The expected total 
density of magnetic moments in the F-ACF (red stars) obtained by multivariable analysis with the 
contributions of the edge state and $\sigma$-dangling bond state.}
  \label{Fig:5}
\end{figure}

Let us discuss the effect of fluorination in the ACFs next. Figure~\ref{Fig:3} shows the C K-edge 
NEXAFS spectra of the fluorinated ACFs at various fluorine concentrations. With an increase in the 
fluorine concentration, the intensity of the $\pi^{\ast}$ peak decreased and a new peak appeared around 290 
eV (denoted by the arrow in Fig.~\ref{Fig:3}), which could be ascribed to the $\sigma^{\ast}$(C-F) \cite{mol2001}. It can be 
understood that the appearance of the new peak is a consequence of the formation of a C-F bond at 
the expense of a $\pi$ bond between the carbon atom attacked by a fluorine atom and the carbon atom 
adjacent to it. Interestingly, upon fluorination, an extra peak (p2) at 284.9 eV was created in 
addition to the edge-state peak (p1) and the $\pi^{\ast}$-conduction band. This peak became pronounced at 
F/C$\sim$0.84, and then it tended to get smaller above that concentration. Accordingly, the spectral 
analysis of the NEXAFS was carried out with three peaks as exhibited in Fig.~\ref{Fig:4}. The energy and 
width of the p2 peak were determined by the spectrum of ACFs at F/C=1.1 with the energy and 
width of peak 2 being 284.9 eV and 0.71 eV, respectively. The energy and peak width of the $\pi^{\ast}$ and 
edge state peaks were fixed to the values of HOPG and the pristine ACFs. Figure~\ref{Fig:5}(c) summarizes 
the integrated intensities of the $\pi^{\ast}$, edge state and p2 peaks as a function of fluorine concentration. 
With the increase in fluorine concentration, the intensity of the $\pi^{\ast}$ and edge state peaks decreased 
monotonically. Peak 2 started appearing at F/C=0.46, and it increased to a maximum at F/C=0.8, 
after which it began to vanish as the fluorine concentration reached the saturated value of F/C$\sim$1.2.

In understanding the behavior of the NEXAFS spectra, we recall the two-stage change in the 
localized spin concentration with a discontinuity around F/C$\sim$0.4 as a function of fluorine 
concentration \cite{jpsj2001}. Figure~\ref{Fig:5}(d) shows the localized spin concentration ($Ns$) 
observed from magnetic 
susceptibility measurements \cite{jpsj2001}. In the range of up to F/C$\sim$0.4, $Ns$ decreased monotonically as 
fluorination proceeded. Above that fluorine concentration, $Ns$ increased to a maximum of around 
F/C$\sim$0.8, after which it decreased toward zero as the fluorine concentration approached the 
saturation concentration of F/C=1.2. This change owes to the stepwise fluorination in the initial and 
successive stages, in which the chemically active edge carbon atoms and the less chemically active 
interior carbon atoms participate \cite{jpc1999}.

In the initial stage (Fig.~\ref{Fig:5}(a)), the termination of edge carbon atoms with fluorine atoms leads 
to the conversion of a graphite sp$^{2}$ bond into an sp$^{3}$ bond at the expense of a $\pi$ bond in the edge 
region. Taking into account the schematic model of the nanographene sheet in ACFs consisting of 
216 carbon atoms, the numbers of the edge carbon atoms connected with two and three carbon 
neighbors in the periphery are given to be 36 and 30, respectively. Carbon atoms at the former edge 
sites form CF$_{2}$ bond and those at the latter sites give CF bonds. The fluorination at the edges is 
eventually completed at ca. (2$\times$36+30)/216 $\sim$47$\%$, which roughly meets the boundary concentration 
of F/C$\sim$0.4. Accordingly, the fluorination of the edge carbon atoms, which makes the conjugated 
$\pi$-electron system shrink, explains the decrease in the intensity of the $\pi^{\ast}$ peak upon the fluorination 
of the edge carbon. The termination with fluorine atoms also affects the edge state. Indeed, 
according to the band structure of graphene nanoribbons calculated by DFT, the flat band at the 
Fermi energy, which is assigned to the edge state, is modulated by fluorination \cite{jpsj2004}. Moreover, 
theoretical calculations indicate that the total magnetic moment for the fluorinated graphene 
nanoribbon, in which the edge on one side is monofluorinated and that on the other side is 
difluorinated, is only 1/3 of that of the hydrogenated nanoribbon \cite{jpsj2004}. This is what we observed in 
the edge-state peak in the NEXAFS spectra as well as in the magnetic behavior observed in the 
susceptibility. Here, we pay careful attention to the effect of fluorination on the magnetic properties 
of ACFs in the range of up to F/C$\sim$0.4. The concentration of localized spin $Ns$ obtained from the 
spin susceptibility $\chi_{S}$ decreases monotonically with fluorination. This means that the termination 
of edge carbon atoms with fluorine atoms kills the localized spins successively. Here, we should 
reconfirm the origin of the localized spin, though the NEXAFS spectrum (p1) suggests it to be 
edge-state spin. In the fluorine concentration region below F/C$\sim$0.4, the negative orbital 
susceptibility is observed for ACFs, which is the signature of the extended conjugated $\pi$-electron 
system [16]. The orbital susceptibility monotonically decreases with fluorination in the fluorine 
concentration region below F/C$\sim$0.4, indicating a decrease in the size of the $\pi$-electron system, in 
accordance with the decrease in the intensity of the $\pi^{\ast}$ peak. Eventually, the localized spins 
observed in this fluorine concentration region were assigned to the edge-state spin of $\pi$-electron 
origin.

After the fluorination is completed at the edge around F/C$\sim$0.4, fluorine atoms begin to attack 
the carbon atoms in the interior of the nangraphene sheet, breaking the $\pi$ bonds (see Fig.~\ref{Fig:5}(b)). The 
decrease in the intensity of the $\pi^{\ast}$ peak is explained by this fluorination process. When a fluorine 
atom is bonded to a carbon atom in the interior, a $\sigma$-dangling bond having a localized spin is created 
at the carbon site adjacent to the carbon atom that is fluorinated. The concentration of the dangling 
bonds becomes maximized when about half of the carbon atoms in the interior of the nanographene 
sheet are singly bonded to fluorine atoms. Again, considering the schematic model of the 
nanographene sheet consisting of 216 carbon atoms, the number of interior carbon atoms is 150. 
Therefore, the total number of fluorine atoms bonded to carbon atoms is 75 (interior site)+102 
(edge site)=177 when half of the carbon atoms in the interior of nanographene sheet are bonded to 
fluorine. Then the fluorine concentration in this situation is F/C=0.82. This estimate coincides with 
the presence of the maximum spin concentration at F/C$\sim$0.8 as shown in Fig.~\ref{Fig:5}(d). The saturated 
fluorine concentration corresponds to the concentration at which all the carbon atoms are 
fluorinated, and is obtained from the schematic model of nanographene as F/C=(150 (interior 
site)+102(edge site))/216$\sim$1.17, which agrees with the observed saturated fluorine concentration of 
F/C=1.2, at which the spin concentration decreased to zero as shown in Fig.~\ref{Fig:5}(d). It is noteworthy 
that the change in the intensity of peak 2 ($\sigma$-dangling bond) tracks the change in the spin 
concentration shown in Fig.~\ref{Fig:5}(d) with fidelity. The good quantitative agreement indicates that the 
$\sigma$-dangling bond observed as peak 2 is responsible for the localized spin observed in the fluorine 
concentration range of 0.4$<$F/C$<$1.2. This assignment is justified from the fact that the orbital 
susceptibility of $\pi$-electron disappears above F/C$\sim$0.4 \cite{jpsj2001}. 
Interestingly, peak 1 still survived even 
at concentrations higher than F/C$\sim$0.4, although its intensity decreased significantly. The 
fluorination process might not be uniform in the ACFs. A part of the edge carbon atoms might not 
be fully terminated with fluorine atoms for some nanographenes sheets, even if fluorine atoms 
begin to attack the carbon atoms in the interior of other nanographenes sheets. Alternatively, the 
existence of the p1 peak might prove the edge-state can coexist even under the condition in which 
the $\pi$-conjugated system in the interior of nanographene sheets is largely destroyed by fluorination.

Here, we briefly comment on the energy and peak width of the edge state and $\sigma$-dangling bond 
state peaks. The peak energy was higher for the $\sigma$-dangling bond state peak (p2: 284.9eV) than that 
for edge state peak (p1: 284.5eV). Furthermore, the peak width of p2 (0.71 eV) was slightly smaller 
than that of p1 (0.82 eV). The peak widths of the p1 and p2 were much smaller than that of the $\pi^{\ast}$ 
peak. These findings can be explained by considering the band dispersion and screening of the core 
hole formed during X-ray absorption. Namely, in contrast to the extended feature of the $\pi^{\ast}$ 
conduction electron state, both the $\sigma$-dangling bond state and the edge state are localized states with 
negligible dispersion \cite{jpsj1996,jpsj1998,jpsj2004}. These features are well reproduced in the observation that the peak 
widths of the $\sigma$-dangling bond state and the edge state peaks are much smaller than that of the $\pi^{\ast}$ 
peak. The experimental finding that the edge state has the width slightly wider than the $\sigma$-dangling 
bond is in good accordance with theoretical indication showing that it spreads along the zigzag edge 
carbon atoms \cite{jpsj2004}. The difference in the electronic features between the edge state and $\sigma$-dangling 
bond state has also been confirmed experimentally by the presence/absence of exchange interaction 
in the edge-state spins/$\sigma$-dangling bond spins \cite{jpsj2001}. The lower energy of the edge-state peak than the 
$\sigma$-dangling bond state peak is due to the strong and weak screening of the core hole for the former 
and latter, respectively; the larger local density of states at carbon atoms of zigzag edges, at which 
the edge states are populated, and the smaller local density of states of the $\sigma$-dangling-bond carbon 
atoms, at which $\pi$-electron are absent \cite{jpc2011}.

Finally, we should note the importance of the combination of magnetic susceptibility and 
NEXAFS analyses in investigating the magnetic properties of F-ACFs, in which two types of 
localized spins coexist; edge state spins and $\sigma$-dangling bond spins. It is difficult to distinguish the 
spin species independently only by the magnetic measurement, whereas NEXAFS provides the 
relative density of states of the edge state and dangling bond states \cite{ass}. Therefore, the combined 
analysis can allow us to estimate the magnetic moments of edge-state spins and $\sigma$-dangling bond 
spins using multivariable analysis. The total density of magnetic moments ($Ns_{i}$) of sample $i$ is given 
as 
\begin{equation}
Ns_{i}=n_{edge,i}\mu_{edge}+n_{\sigma,i}\mu_{\sigma}
\end{equation}
where the $n_{edge,i}$ ($n_{\sigma,i}$) and  $\mu_{edge}$ ($\mu_{\sigma}$ ) are the density and the 
magnetic moment, respectively, of the edge state ($\sigma$-dangling bond state). 
The $\mu_{edge}$ and $\mu_{\sigma}$ were 
obtained by minimizing 

\begin{equation}
S=\displaystyle \sum_{i=1}^k (Ns_{i}-n_{edge,i}\mu_{edge}-n_{\sigma,i}\mu_{\sigma})^{2}
\end{equation}

where $k$ was the number of samples.  
$\mu_{edge}$ and $\mu_{\sigma}$ were, thus, given by 

\begin{equation}
\mu_{edge}=\frac{\sum n_{edge,i}Ns_{i}\sum (n_{\sigma,i})^2-\sum n_{\sigma,i}Ns_{i}\sum n_{edge,i}n_{\sigma,i}}
{\sum (n_{\sigma,i})^{2}\sum (n_{edge,i})^{2}-(\sum n_{\sigma,i}n_{edge,i})^{2}}
\end{equation}

and 

\begin{equation}
\mu_{\sigma}=\frac{\sum n_{\sigma,i}Ns_{i}\sum (n_{edge,i})^2-\sum n_{edge,i}Ns_{i}\sum n_{edge,i}n_{\sigma,i}}
{\sum (n_{\sigma,i})^{2}\sum (n_{edge,i})^{2}-(\sum n_{\sigma,i}n_{edge,i})^{2}}
\end{equation}

respectively. The relative density of the edge state ($n_{edge,i}$) and dangling bond state ($n_{\sigma,i}$) were 
evaluated by the integrated peak intensity of the corresponding peaks in the NEXAFS spectrum of 
each fluorinated sample $i$. Using $Ns_{i}$ obtained by the SQUID measurement, the ratio of the magnetic 
moments of the edge state to dangling bond state was determined to be $\mu_{edge} : \mu_{\sigma} \sim 0.7(\pm 0.1):1.0$ as 
shown in Fig.~\ref{Fig:5}(d). The magnetic moment of the edge state was smaller than that of the dangling 
bond state, in good agreement with theoretical prediction which gives an estimate of the magnetic 
moments of the edge state and the dangling bond state as $\sim$0.5$\mu_{B}$ and 1$\mu_{B}$, respectively \cite{prb2007}. The 
fractional magnetic moment of the edge state spin is associated with the itinerant nature of the 
edge-state electrons, whereas the localized nature of the dangling bond is responsible for 
non-fractional moment of 1$\mu_{B}$.

\section{CONCLUSIONS}
\label{sec5}
We investigated the electronic structure of ACFs and fluorinated ACFs using NEXAFS whose 
results were complementary to the magnetic measurement results of the same samples. From the 
NEXAFS spectra, the edge state, which is proved to have localized spins by the magnetic 
measurements, is suggested to exist below the $\pi^{\ast}$ peak of ACFs in NEXAFS. Upon fluorination, the 
intensity of the edge state peak and $\pi^{\ast}$ peak decreased monotonically. An extra peak appeared for 
ACF in the region of F/C=0.4$\sim$1. The intensity of this extra peak closely correlated with the fluorine 
concentration dependence of the localized spin concentration, which confirmed the existence of the 
magnetic $\sigma$-dangling bond state for the fluorinated ACFs. The experimental findings obtained by 
the combination of electronic (NEXAFS) and magnetic techniques importantly confirm the 
coexistence of the magnetic $\pi$-edge state and $\sigma$-dangling bond state, the former and latter of which 
have itinerant and localized nature, respectively. 

\section{ACKNOWLEDGMENTS}
The present work was performed under the approval of PF-PAC (No. 2009G022, 2010G036). 
The authors are grateful for the financial support from MEXT (the Grant-in-Aid for Scientific 
Research No. 20001006). We would like to express our sincere gratitude to Prof. K. Terakura at 
JAIST for the stimulating discussion.


\begin{thebibliography}{0}
\bibitem{ass}We assume that the center of the edge state or dangling bond state is positioned at the Fermi 
level, that is, the density of states for the occupied state is equal to that for the unoccupied state. 
We also assume that the transition moments from C 1s to $\pi^{\ast}$, edge state, and dangling bond 
states are equal to each other.
\end{thebibliography}
\end{document}